\documentclass[a4paper]{jpconf}
\usepackage{graphicx}

\begin{document}
\title{Jets and Underlying Events at LHC Energies}
\author{A.~G.~Ag\'ocs$^{1,2}$, G.~G.~Barnaf\"oldi$^1$, P.~L\'evai$^1$}
\address{$^1$MTA KFKI RMKI, Konkoly-Thege Mikl\'os \'ut 29-33, H-1121 Budapest, Hungary}
\address{$^2$E\"otv\"os University, P\'azm\'any P\'eter s\'et\'any 1/A, H-1117 Budapest, Hungary}

\ead{agocs@rmki.kfki.hu}

\begin{abstract}
  Jet-matter interaction remains a central question and a theoretical
  challenge in heavy-ion physics and might become important in
  high-multiplicity events in proton-proton collisions at LHC energies.
  Full jet measurement at LHC offer the proper tool to investigate energy
  loss process and fragmentation of hard parton in the medium. Since jet
  reconstruction will be constrained to small cone sizes, then study of the
  connection between jets and surrounding environment provides a further
  possibility to extend our exploration. We study jets at $\sqrt{s}=14$ TeV
  and $pp$ collisions at $\sqrt{s}=7$ TeV. We analyze the flavor components
  in jet-like environments. We introduce a definition for surrounding
  cones/belts and investigate flavor dependence and correlation of
  different hadron species produced in jets.  Here, we focus on
  proton-triggered correlations. Our analysis can be extended for heavy ion
  collisions.
\end{abstract}

\section{Introduction}

The state-of-the-art particle detectors and very-high energies reached at the 
Large Hadron Collider (LHC) opened a new window to create and investigate high 
momentum particle showers in hadron-hadron collisions. 
These showers originate from color partons and they are identified as 
{\sl jets}. Jets were discovered in electron-positron ($e^-e^+$) collisions, 
where it was assumed: partons are travelling in QCD vacuum. 
However, in proton-proton  and heavy-ion collisions partons are propagating
in a color medium. The deconfined medium modifies the parton properties e.g. induces 
gluon radiation and parton energy loss~\cite{GLV}. 
This phenomena has been seen and investigated in details
in heavy-ion collisions at RHIC energies~\cite{Levai:2001dc,Vitev:2009}. 

High-multiplicity proton-proton events may behave similarly displaying intense
interaction between jets and the surrounding 'matter'. Thus, study of jet energy 
loss and jet-matter interaction can be accomplished with the proper 
determination of the jets and the remaining background, the so called 
{\sl underlying events}.  

Generally speaking, underlying events (UE) contain particles originating
from partons outside identified jet(s). The first 'standard' definition of 
UE was given by the CDF Collaboration at Fermilab~\cite{CDFUE}. 
In an earlier work we introduced {\sl surrounding belts} (SB) around the
cones of identified jets~\cite{Agocs:2009}. 
Here, we focused on properties of near and away side jets, and their 
surrounding regimes, especially hadron contents and correlations. 

We display our recent results on the generalized definition of the
underlying event. We analyzed mean-$p_T$ vs. multiplicity for the newly
defined areas in case of $\sqrt{s}=14$ TeV proton-proton collisions.  We
performed PYTHIA simulations for $\sqrt{s}=7$ TeV $pp$ collisions. We used
PYTHIA 6.4~\cite{Pythia6} with ATLAS-CSC tune~\cite{atlas-csc} for both
cases and generated $\sim 20$ M minimum bias events. Furthermore, we have
studied proton-triggered correlations in surrounding belts characterized by
various radii and thicknesses at $\sqrt{s}=7$ TeV.

\section{Mean-$p_T$ vs. multiplicity for the generalized UE}
\label{Angular-Section}

Transition area between jet cones and underlying event carries information about
jet-matter interaction. The investigation of this area demands proper definitions
and solid basis for quantitative analysis.
  
\begin{figure}
\begin{center}
\includegraphics[width=75mm,height=65mm]{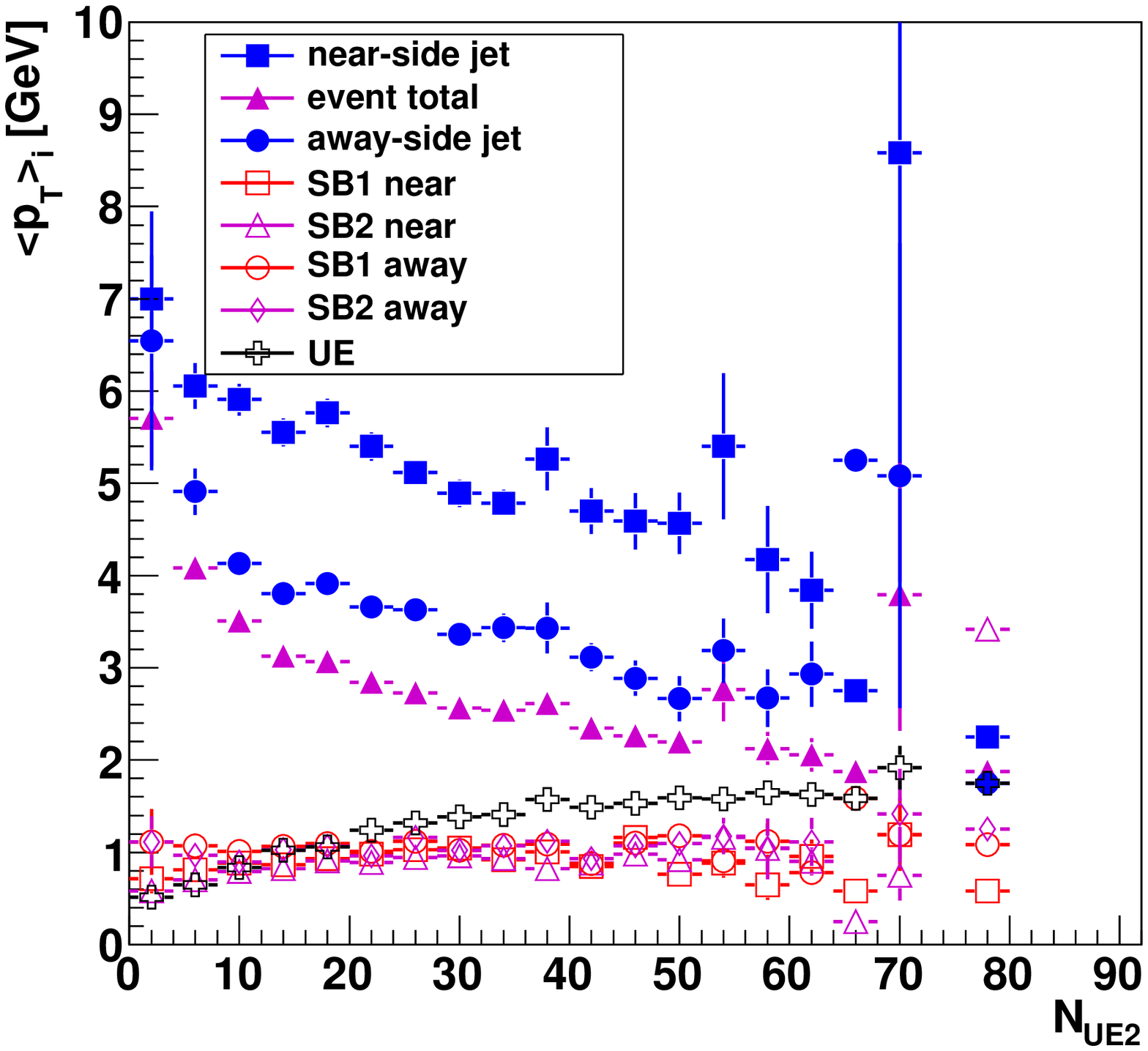}
\includegraphics[width=75mm,height=65mm]{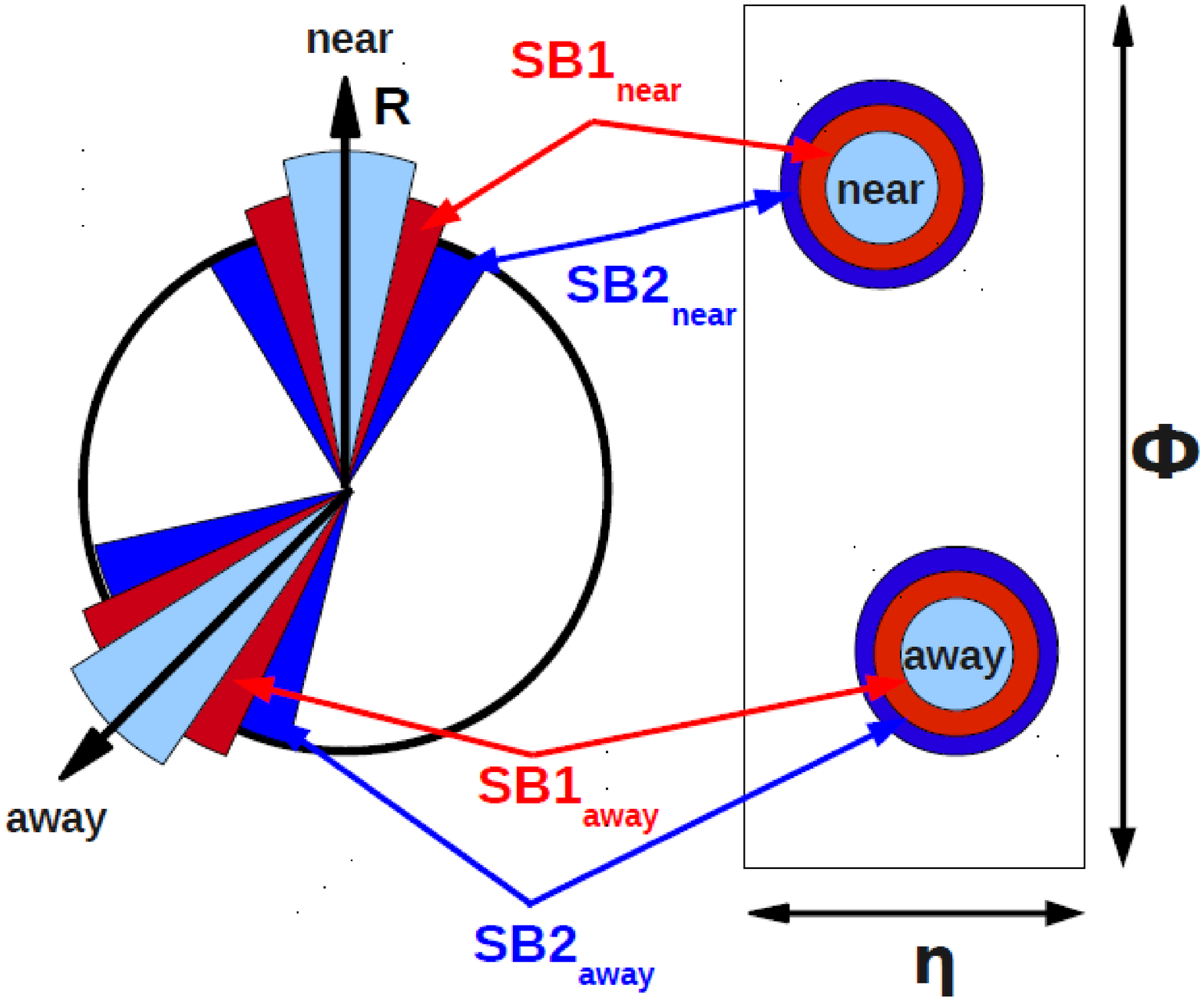}
\end{center}
\caption{\label{ang-corr} {\sl Left panel:} mean-$p_T$ vs. UE multiplicity outside 
surrounding cones in $pp$ collisions at $14$ TeV. {\sl Right panel:} illustration of 
'inner-belt' (SB1) and 'outer-belt' (SB2). (Color online.)}
\end{figure}

First we considered the mean transverse momenta vs. multiplicity in 
$14$ TeV proton-proton collisions. The 
newly defined areas based on {\sl right panel} of Fig.~\ref{ang-corr}. 
Here, the main areas are highlighted: 'near' and 'away' refers for the 
the near- and away-side jets respectively. We marked the inner surrounding 
belt with 'SB1' and the outer 'SB2' for both near and away side jets. 
Based on this definition, the underlying event is everything outside the 
outer belts.  

Hadron multiplicities can be given in each region however, for reference we
used $N_{UE2}$, which is the generalized underlying event area -- outside
both outer surrounding belts. We plotted the mean transverse ($\langle p_T
\rangle$) momenta of above areas as the function of $N_{UE2}$ on the {\sl
  left panel} of Fig.~\ref{ang-corr}. The ($\langle p_T \rangle$ for near-
and away-side jet is drawn by {\sl full squares} and {\sl full disks}
respectively. Mean-$p_T$ for surrounding belts are plotted with {\sl open
  squares} and {\sl open triangles} for inner belts: 'SB1$_{near}$' and
'SB1$_{away}$'. Open circles and diamonds denote results on outer belts:
'SB2$_{near}$' and 'SB2$_{away}$', respectively.  {\sl Full triangles}
display the $\langle p_T \rangle$ for the total event.  Finally, we indicated
$\langle p_T \rangle$ for the underlying event with {\sl stars}.

As we plotted on Fig.~\ref{ang-corr} jet-like events has high-$\langle p_T
\rangle$ values, with low multiplicity, due to the produced jets in the
$14$ TeV $pp$ collisions. Mean $p_T$s for these jet-like events are falling
quickly as the $N_{UE2}$ multiplicity is increasing. In parallel, the
$\langle p_T \rangle$ for the total event has a similar structure.
Surrounding belts have almost the flat value $\langle p_T \rangle = 1$
GeV/c, practically at every $N_{UE2}$ value.  Testing the underlying event,
$\langle p_T \rangle _{UE} $ is increasing up to $\sim 2$ GeV/c, indicating
a 'mini-jet' like structure at high multiplicities is starting to play the
role, as we pointed out in Ref.~\cite{Agocs:2009}.

\newpage
\section{Geometrical setup for the analysis}
\label{Geometry}

The aim of our study is to map differences in surrounding belts, 
connected to near-side and away-side correlation peaks. Such differences 
carry information about hadronization processes and/or possible jet-matter 
interactions. The wanted differences can be amplified displaying the ratio of 
hadron yields in the different belts: SB1$_{near}$/SB1$_{away}$ and 
SB2$_{near}$/SB2$_{away}$. 

Following the illustration of Fig.~\ref{ang-corr}, we define 3 different 
'geometrical' sets to be analyzed: 

\begin{description}
\item[CDF-set]: $\Delta \phi_{near} = 0^{\circ} \pm 60^{\circ}$;  
$\Delta \phi_{away} = 180^{\circ} \pm 60^{\circ}$; 
$\Delta \phi_{SB1} = \Delta \phi_{SB2} = \pm 6 ^{\circ}$.  

\item[$R$-set]: $\Delta \phi_{near} = 0^{\circ} \pm 30^{\circ}$;  
$\Delta \phi_{away} = 180^{\circ} \pm 30^{\circ}$; 
$\Delta \phi_{SB1} = \Delta \phi_{SB2} = \pm 6 ^{\circ}$.  

\item[$\sigma$-set]: $\Delta \phi_{near} = 0^{\circ} \pm \sigma_{near}/2$;  
$\Delta \phi_{away} = 180^{\circ} \pm \sigma_{away}/2 $; 
$\Delta \phi_{SB1} = \Delta \phi_{SB2} = \pm 6 ^{\circ}$.  

\end{description}

The CDF-set is motivated by the original definition from CDF 
Collaboration~\cite{CDFUE}. The $R$-set is 
motivated by a smaller, jet-cone-like size. 
The $\sigma$-set
is determined from PYTHIA simulations on correlations at $7$ TeV at 
corresponding trigger and associated momentum regions. Here we will 
use $\sigma_{near}=16.0 ^{\circ}$ and $\sigma_{away}=19.5 ^{\circ}$ displayed on 
Fig.~\ref{ang-corr}. 

Comparing the above selections, the area of the 
underlying event is the smallest in the case of CDF-set, and it is the largest 
for $\sigma$-set. The widths of the surrounding belts ($6^{\circ}$) is a result of 
an optimalization to have proper statistics for analysis and comparison.

\section{Correlation studies with charged hadron triggers}
\label{UE-Section}

We have calculated the wanted ratios of
SB1$_{near}$/SB1$_{away}$ and
SB2$_{near}$/SB2$_{away}$ for all possible charged hadron triggers, namely
choosing one of the identified hadrons ($\pi^+$, $\pi^-$, $K^+$, $K^-$, 
$p$, $\bar{p}$) as trigger particle and extracting angular correlation 
with all other identified hadron species.
\begin{figure}[h]
\begin{center}
\includegraphics[width=160mm,height=90mm]{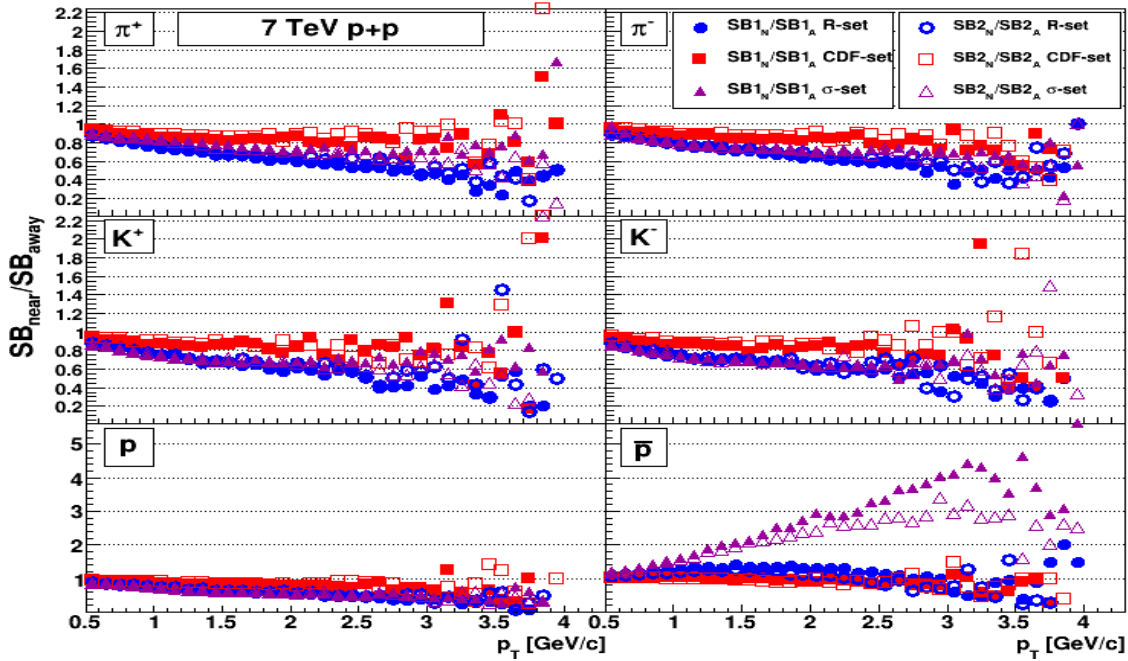}
\end{center}
\caption{\label{7T-p-trigger} Differences in the proton-triggered 
correlation functions, 
enhanced by ratios of surrounding belts, 
SB1$_{near}$/SB1$_{away}$ and SB2$_{near}$/SB2$_{away}$ as a function of 
transverse momentum of the associated hadrons, $p_{T,assoc}$. See text for 
details on legend. (Color online.)  }
\end{figure}

Fig.~\ref{7T-p-trigger} displays our results extracted from PYTHIA simulations
for $pp$ at $7$ TeV, using proton trigger in the momentum window defined 
earlier. 
Inner belt ratios, SB1$_{near}$/SB1$_{away}$ are marked by 
{\sl full squares} for 'CDF-set', {\sl full disks} for  
'$R$-set', and {\sl full triangles} for '$\sigma$-set'. 
Outer belt ratios, SB2$_{near}$/SB2$_{away}$ are marked by 
{\sl open squares} for 'CDF-set', {\sl open circles} for  
'$R$-set', and {\sl open triangles} for '$\sigma$-set'. 
Further results for other identified hadron triggers can be visualized
similarly, but we want to focus on proton trigger.

Results from 'CDF-set', indicated by {\sl full and open squares} are close 
to be unity (no effect)
in all correlations. This means we are testing homogeneous UE region far 
from correlation peaks.    

Results from '$R$-set', indicated by {\sl full and open circles} go below 
unity for most of the hadrons (except antiprotons), because of the expected
momentum conservation for leading hadrons. Antiprotons behave differently due
to a strong correlation in baryon-antibaryon production closer to the 
correlation peaks. This effect is based on baryon number conservation. A 
slight difference can be already seen between the 
ratios from inner belt and outer belt.  

In the case of '$\sigma$-set' ({\sl full and open triangles}) we are closer
to the correlational peaks, but the results are very similar to the '$R$-set',
except the deviation in proton-antiprotons correlation is even more strong.
Thus, we are really in the fragmentation region with very strong 
baryon-antibaryon correlation. A slight difference between the inner belt 
and outer belt can be seen, also.  

Taking $K^+$ trigger we will see strong  $K^+ - K^-$ correlation
driven by strangeness conservation. The $K^+-$proton and $K^+-$antiproton
correlations are just similar to e.g. $K^+ - \pi^+$ ones.

\section{Conclusions}

We have investigated identified hadron correlations in $pp$ collision at
$7$ TeV, by means of PYTHIA event generator. We introduced different 
geometrical setups. We have found strong proton-antiproton correlation, based 
on baryon number conservation and strong $K^+ - K^-$ correlation driven by 
strangeness conservation. Introducing different surrounding belts into 
the analysis further details were revealed. These correlations should 
be extracted from real data. The ALICE HMPID and the planned VHMPID 
detectors can play an important role in these explorations~\cite{HMPID,VHMPID}.   

\ack

This work was supported by Hungarian OTKA NK77816, PD73596 and
E\"otv\"os University. One of the authors (GGB) is also thanks to J\'anos Bolyai 
Research Scholarship of the HAS.


\end{document}